\documentstyle[preprint,revtex]{aps} 
\begin{document}
\draft
\preprint{
\begin{flushright}
LAEFF--93/006\\
April 1993
\end{flushright}
}
\begin{title}
Randomness and Irreversibility\\
in \\
Quantum Field Theory.\\
\end{title}
\author{J.\ P\'erez--Mercader\cite{AAAuth}}
\begin{instit}
Laboratorio de Astrof\'{\i}sica Espacial y F\'{\i}sica
Fundamental\\
Apartado 50727\\
28080 Madrid
\end{instit}
\vspace{-.5in}
\begin{abstract}
Quantum fluctuations, through quantum corrections, have the
potential to lead to irreversibility in quantum field theory. We
consider the virtual ``charge" distribution generated by quantum
corrections in the leading log, short range approximation, and adopt
for it a statistical interpretation. This virtual charge density has
fractal structure, and it is seen that, independently of whether the
theory is or is not asymptotically free, it describes a system where
the equilibrium state is at its classical limit ($\hbar \rightarrow
0$). We also present a simple analysis of how diffusion of the charge
density proceeds as a function of the distance at which the
system is probed.
\end{abstract}
\vspace{-.75in}
\begin{center}
\bf {\sl Submitted to Physical Review Letters}
\end{center}
\vspace{.25in}
\narrowtext

Quantum fluctuations are perhaps the most fundamental feature of
quantum field theory, affecting fields, their sources and the vacuum in
which they evolve. Induced by  the fact  that  quantum  fluctuations
are to a certain  extent  random \cite{bell}, we can expect them to
have an impact on questions of irreversibility.  Namely,  since
$true$  randomness impairs one's  ability  to  carry out  an $exact$
reconstruction of  the past  history of  the system, one can expect
some irreversibility at the  microscopic (quantum field) level because
of the presence of quantum fluctuations in this domain. On the other
hand, as  is well known, quantum  fluctuations through  their breaking
of canonical scale--invariance, induce non--trivial scaling in physical
parameters and turn them into scale dependent, effective parameters,
whose scale dependence is  described by the appropriate
renormalization group equations (RGE) \cite{gellmannlow}.

There is then a relationship between irreversibility and non--trivial
renormalization group scaling.

In fact, in quantum field theory one associates with the
quantum--corrected interaction energy a charge density which results
from taking into account (through quantum corrections) not only simple
two body processes with a precise and fixed impact parameter, but
$also$ the effects of many body interactions with impact parameter
similar or smaller than the distance scale at which we probe the
system. The measured physical potential then is subject to
fluctuations, and it is impossible to specify the $exact$ dependence
of the potential on the individual components of the virtual charge
density. The effect of quantum fluctuations on the scaling properties
of the potential and the associated effective charge density, are then
a manifestation of statistical indeterminacy.

To handle questions relating to the effective charge density we need
to introduce some statistical notions into the quantum field theory.
The fluctuations in the potential are related to the ones in the
charge density through the Poisson equation satisfied by the potential
and the charge density. Under very general circumstances the induced
density can be interpreted as a probability density \cite{doob}, and
the needed statistical framework just unfolds in front of our eyes.

We will compute the effective interaction energy and derive from it the
charge density. The effective interaction energy is obtained by
solving its RGE. Since energy has canonical dimension of inverse time
and no anomalous dimension, its RGE \cite{rothe} has the solution,

\begin{equation}
V(\lambda r_0,g_0,a)=\lambda ^{-1}V(r_0,\bar{ g}_0(\lambda ),a)\,
{}.
\label{1}
\end{equation}

\noindent
Here $V$ is the interaction energy, $r_0$ is the distance between the
interacting sources, and $a$ is a reference distance. The quantity
$\lambda$ is the scale parameter, $\bar{g}_0(\lambda )$ is the
effective coupling and satisfies the RGE $\lambda \partial \bar{g}_0 /
\partial \lambda = - \beta (\bar{g}_0)$. For a massless  mediating
field (such as photons or gravitons), the effective  interaction
energy  for two point particles separated by a distance $a$ is given by
$V(a,g_0,a)=C\frac{g_0^2}{4 \pi a} $. Computing to 1--loop order,
where $\beta=-\beta_0 g_0^3$,  and choosing $\lambda =r/a$, we get
(for $r$ close to $a$ and after exponentiating)

\begin{equation}
V(r,g_0,a) =C\frac{g_0^2}{4 \pi}a^{-\sigma} r^{-1+\sigma} \, .
\label{8/1}
\end{equation}

\noindent
$\sigma$ is related to the $\beta$--function for the coupling $g_0$,
and to one--loop is given by $\sigma=-2 \beta_0 g_0^2$ and we have
taken it to  be a constant.\footnote{Strictly speaking, $\beta$ is a
function of the distance at which the system is probed. This can be
made explicit by computing it in a mass dependent subtraction
procedure \cite{gandp}. However, it will be sufficient for our
$present$ purposes to take it to be a constant.} \footnote{Note that
one can also arrive at the result shown in eq. (\ref{8/1}) by direct
solution of the RGE for the propagator of the particle mediating the
force.} For  the case of QED one recognizes  here  the first term  in
the short  distance  expansion  of  the   famous  Uehling  potential;
in QCD one recognizes the expression for the interquark potential
\cite{fischler}.

Potential theory guarantees that eq. (\ref{8/1}) satisfies a Poisson
equation whose right hand side is proportional to the ``charge"
density dictated by quantum corrections. This charge density can be
understood \cite{doob} as a ``probability density", and used to derive
information on the physics of the virtual cloud as a many body
(statistical) system. For our isotropic potential, the density is

\begin{equation}
\rho (r)=A' \ \ r^{-3+\sigma}
\label{10-[1]}
\end{equation}

\noindent
where $A'$ is a constant. For $\rho (r)$ to be a probability density,
we need that it be a positive and integrable function on its support.
How this is implemented depends on the sign $and$ size of $\sigma$
(see below).

{}From eq. (\ref{10-[1]}) we see that $\rho (r)$ is the solution to the
functional equation

\begin{equation}
\rho (\lambda r)=\lambda ^\beta \rho (r)
\label{functional}
\end{equation}

\noindent
with $\beta= -3+\sigma$. Thus $\rho(r)$ describes a fractal
distribution of charge embedded in  a 3--dimensional configuration
space; the Hausdorff (or fractal) dimension is $d_f=+\sigma$. This is
not surprising, since $\sigma$ has its origin in the $deviation$ from
$canonical$ scaling of the effective coupling due to quantum
fluctuations. Furthermore,  since $\sigma$ changes as we change the
size of the domain  on which  we probe the  virtual cloud,
\footnote{Because it is related to the number of degrees of freedom
that contribute at the scale on which we study the system.} it turns
out that the Hausdorff dimension also changes: in truth we are  then
dealing with a multifractal. For QCD and Quantum Gravity in their
asymptotically free regime, $\sigma$ is positive. In QED and other
non--asymptotically free theories, $\sigma$ is negative.

Normalizability of the probability density, leads to

\begin{equation}
\rho=A r^{-3+\sigma}
\label{density}
\end{equation}

\noindent
with $A={\sigma  \over {4\pi }}R_0^{-\sigma }$ for asymptotically free
theories, and $A=-{\sigma  \over {4\pi }}r_0^{-\sigma }$ when $\sigma
< 0$. Here $R_0$ denotes an IR--cutoff, necessary in the case of
positive $\sigma$, in order to ensure the finiteness of the probability
distribution. For non--asymptotically free theories $r_0$ is an
UV--cutoff necessary because of the same reason. The IR--cutoff can be
identified in QCD with, e. g., a typical hadronic size; the UV--cutoff
$r_0$ can be identified in QED with the Compton wavelength for the
electron, or for the heaviest degree of freedom considered. In either
class of theories the resulting probability density is of the Pareto
type \cite{schroeder}; this is a natural consequence of the
renormalization group origin for $\rho (r)$, which is ultimately
responsible for the functional equation in (\ref{functional}) and the
associated scaling: scaling leads to Levy--type distributions, which
turn Pareto in some limit \cite{montroll}. Notice also that since
$\sigma$ vanishes in the classical ($\hbar \rightarrow 0$) limit, then
both $probability$ densities go to zero in this limit.

One may now introduce a quantity

\begin{equation}
S=-k\int_{}^{} {d^3}\vec r\rho (r)\log \left[ {C_N \rho (r)}
\right] \label{entropy}
\end{equation}

\noindent
with the properties of a fine--grained entropy, and study the
``equilibrium configurations" that these distributions support.
We will $identify$ the configurations (as in information theory
and thermodynamics) corresponding to the maximum entropy with
the equilibrium configurations of the system. In other words, we
will assume that this fine--grained entropy reveals a preferred
``direction"  for stability in the evolution of the quantum
field system. In eq. (\ref{entropy}), the integral extends
over the full support of the variable $r$. The constant $C_N$
needs to be introduced because $\rho (r)$ is a dimensionful
quantity, and we do not have the equivalent of a Nernst theorem
to set a reference valid for $all$ physical systems.

When $\sigma$ is positive

\begin{equation}
S^{\left( {\sigma >0} \right)}=1-{3 \over \sigma }-\log {\sigma  \over
{4\pi }}C_N+3 \log R_0 \,\, ,
\end{equation}

\noindent
and for negative $\sigma$ one gets,

\begin{equation}
S^{\left( {\sigma < 0} \right)}=1-{3 \over \sigma }-\log {-\sigma  \over
{4\pi }}C_N+3 \log r_0  \,\, .
\end{equation}

The ``entropy constant", $C_N$, \cite{fermi}, may be chosen so as to
eliminate the dependence of the entropy on the cutoff. The
``Nernst--volume" corresponds to the extreme volume that can be
physically reachable. These entropies are shown in figs. 1 and 2.

For non--asymptotically free theories, the maximum of the entropy
happens when $\sigma$ goes to 0 from the left. This means either the
classical limit, in the sense that $\hbar \rightarrow 0$, or that one
probes the system at distances {\it so large} that no quantum
fluctuation contributes to $\sigma$. It corresponds to a first order
phase transition.\footnote{Naturally, this also happens when the
interaction is turned off and $g_0^2 \sim 0$.}

For asymptotically free theories the situation is different. The
entropy has a maximum at $\sigma =3$ and a discontinuity at $\sigma
=0$. The maximum value of the entropy occurs when $\sigma$ is close to
the non--perturbative regime, where we can not trust the
approximations made in this paper; in perturbative regimes, and since
$matter$ fluctuations contribute negatively to $\sigma$, the most
stable state corresponds to a configuration (or system size) where
matter states cannot be excited and do not contribute. Because of the
decoupling theorem \cite{decoupling}, this occurs at the {\it
maximum--possible size} of the system: that is, at its IR--limit!

Thus, for both, asymptotically--free and
non--asympto\-ti\-cally--free  theories one sees that {\it the maximum
of the entropy is attained at the classical limit}.

The randomness in the virtual cloud will affect the way into which it
self--organizes and spreads into space--time. The fractal nature of
the probability density leads one to expect some form of fractal
brownian motion to be present. The parameter controlling this is
$\sigma$, as may be seen by examining some basic properties of the
coefficient of diffusion derived from $\rho (r)$.

Taking the collision probability among the components of the virtual
cloud to be $\rho(r)$, the virtual charges will execute random walks
with individual steps controlled by $\rho(r)$; the net effect of these
processes is the diffusion of the virtual cloud outside of a ball  of
radius $R$, and within which the original interaction is operative.

As is well known \cite{chandrasekar}, after $N$--steps, the
probability that a ``particle"  be at a position between $\vec r$ and
$\vec r + d\vec r$,
 is given by

\begin{equation}
W(r)d^3 \vec r=\left[ {4\pi Dt} \right]^{-3/2}\exp \left[ {-\left|
{\vec r} \right|^2/\left( {4Dt} \right)} \right]d^3\vec r
\label{17/1}
\end{equation}

\noindent
where $D$ is the coefficient of diffusion,

$$
D={n \over 6}\left\langle {\vec r^2} \right\rangle \,\, ,
$$

\noindent
$n$ is the number  of  steps per unit time and $N=nt$ is the total
number of  steps.  $\left\langle {\vec r^2} \right\rangle$ is the
second moment  of the probability  density $\rho (r)$.  When $D=0$,
there is no diffusion.

{}From a radial distance $s_0$ to a radial distance $s_1 > s_0$, the
second moment of $\rho (r)$ is

\begin{equation}
\left\langle {\vec r^2} \right\rangle=\frac{4 \pi}
{2+\sigma}  A
 \left(
s_1^{2 + \sigma} -s_0^{2+\sigma}
\right) \,\, ,
\end{equation}

\noindent
and therefore, in some cases, the coefficient of diffusion $diverges$
(naively) as $s$ goes to infinity. This was to be expected,  since our
probability distribution is  the limit of a Levy--type distribution,
and hence  its moments diverge.

When $\sigma > 0$, we can take $s_1 =R_0$ and set $s_0=s$, an
unspecified but $fixed$ distance. The coefficient of diffusion is
given by

\begin{equation}
D^{(\sigma > 0)}=\frac{n}{6} \frac{\sigma}{2+\sigma} s^2
\left[ \left( \frac{R_0}{s} \right) ^2-
\left( \frac{s}{R_0} \right) ^\sigma
\right] .
\label{D+}
\end{equation}

\noindent
This has the property that it vanishes both when $\sigma \rightarrow
0$ (e.g., as in the classical limit) and when $s \rightarrow R_0$. In
the limit when $s/R_0 << 1$, $D^{(\sigma > 0)} \rightarrow \frac{n}{6}
\frac{\sigma}{2+ \sigma} R_0^2$, and we see the divergence associated
with Levy--Pareto distributions. A plot of $D^{(\sigma > 0)}$ is shown
in fig. 3. From the above, it follows that in the classical $and$
infrared limits there is no diffusion.

In non--asymptotically free theories, we take $s_0=r_0$ (the
UV--cutoff) and now leave $s$ unspecified. Then,

\begin{equation}
D^{(\sigma < 0)}=\frac{n}{6} \frac{-\sigma}{2+\sigma} s^2
\left[
\left(
\frac{s}{r_0}
\right)^{\sigma}
-
\left(
\frac{r_0}{s}
\right)^2
\right].
\label{D-}
\end{equation}

\noindent
As before, $\lim_{\sigma \rightarrow 0} D^{(\sigma < 0)} =0$; also,
when $s$ approaches the UV--cutoff, $D^{(\sigma < 0)}$ goes to zero.
Hence no diffusion of the cloud takes place at either the classical
limit, or at the UV--cutoff. Figure 4 shows a plot of $D^{(\sigma <
0)}$. When the limit of $s$ well into the IR is taken, we see that
there are two possible regimes, depending on whether $|\sigma| < 2$ or
$|\sigma| > 2$. When $|\sigma| < 2 $, $D^{(\sigma < 0)}$ diverges as
$\left( \frac{s}{r_0} \right) ^{2-|\sigma|}$, whereas when $|\sigma| >
2$, $D^{(\sigma < 0)}$ tends to the limit $\frac{n}{6} r_0^2$; the
first case is the signature of a Levy--Pareto distribution, while the
second indicates the presence of a classical brownian motion process.
Notice also that for $|\sigma| \leq 1.5 $, at a given value of
$\sigma$, there always is a maximum value of the diffusion coefficient
that happens when $s/r_0$ falls on the apex of the surface plotted in
Figure 4; at these values the virtual cloud diffuses at the fastest
rate.

In summary, we have exploited the randomness associated with quantum
fluctuations, and applied a statistical interpretation to the virtual
charge they generate. This has interesting and apparently deep
consequences which stem from the scaling properties of the interaction
energy. In particular, fractal behavior in a form related to the
Levy--Pareto form of the charge density, leads to systems where the
maximum entropy is associated with the classical limit of the quantum
system; in other words, the classical regime is the most stable, and
the one preferred by the system. As is the case for other many body
systems, diffusion is a familiar mechanism for the approach to
equilibrium (although perhaps not the only one). In concordance to
what follows from studying the entropy, the virtual cloud diffuses
with a coefficient of diffusion proportional to $\hbar$, and thus
diffusion of the cloud stops as the classical limit is reached.

We see that, under very general conditions, irreversibility is
contained in quantum field systems, and that this irreversibility is
encountered as the system size grows. Furthermore, the quantum field
system tends to ``relax" into a classical system.

The ideas presented \cite{self} here may also find application in the
study of intermittency, turbulence, phase transitions, multiplarticle
physics and other complex phenomena in quantum field theory and the
early universe.

\noindent

\acknowledgements
I have benefitted from discussions with R. Blanckenbecler, A.
Carro, T. Goldman, A. Gonz\'alez--Arroyo, S. Habib, J. Hartle, S.
Lloyd, M. M. Nieto, E. Trillas and G. West.

\figure{Entropy for a generic asymptotically free theory. See
text for notation.}

\figure{Entropy for a generic non-asymptotically free theory.
See text for notation.}

\figure{The diffusion coefficient as a function of $s/R_0$ and
$\sigma$ for an asymptotically free theory. Here $s$ is a fixed
distance representing the maximum size on which the system is
studied, subject to the constraint that it be smaller than the
IR--cutoff $R_0$.}

\figure{The diffusion coefficient as a function of $s/r_0$ and
$\sigma$ for a non--asymptotically free theory. Here $s$ is a fixed
distance representing the maximum size on which the system is
studied. $r_0$ represents the UV--cutoff.}

\end{document}